\documentclass[12pt]{article}
\usepackage{epsfig}
\usepackage{amsfonts}
\usepackage{amsmath}
\usepackage{amssymb}

\usepackage{color} 
  \newcommand{\beq}{\begin{equation}}
  \newcommand{\eeq}{\end{equation}} 
  \def\nuc#1#2{\relax\ifmmode{}^{#1}{\protect\text{#2}}\else${}^{#1}$#2\fi}
  \def\itnuc#1#2{\setbox\@tempboxa=\hbox{\scriptsize\it #1}
    \def\@tempa{{}^{\box\@tempboxa}\!\protect\text{\it #2}}\relax
    \ifmmode \@tempa \else $\@tempa$\fi}
  
\newcommand{\dslash}[1]{#1 \llap{/\kern-0.5pt}}
\newcommand{\Dslash}[1]{#1 \llap{/\kern+1.2pt}}
\newcommand{\DDslash}[1]{#1 \llap{/\kern+2.3pt}}
\newcommand{\dslashh}[1]{#1 \llap{/\kern+1pt}}

\def\Mlo{M_\mrm{lo}}
\def\Mhi{M_\mrm{hi}}

\newcommand{\mrm}[1]{\mathrm{#1}}

\def\bdm{\begin{displaymath}}
\def\edm{\end{displaymath}}

\begin{document}

\begin{titlepage}

\vspace*{1.5cm}

\begin{center}
{\Large\bf Nuclear Effective Field Theories:}
\\
\vspace{0.3cm}
{\Large\bf Reverberations of the early days} 
\\

\vspace{2.0cm}

{\large \bf U. van Kolck}

\vspace{0.5cm}
{
{\it Universit\'e Paris-Saclay, CNRS/IN2P3, IJCLab,
\\
91405 Orsay, France}
\\
and
\\
{\it Department of Physics, University of Arizona,
\\
Tucson, AZ 85721, USA}
}

\vspace{1cm}

\today

\end{center}

\vspace{1.5cm}

\begin{abstract}
Effective field theories 
are recognized nowadays as 
{\it the} framework to describe low-energy nuclear structure
and reactions consistently with
the underlying theory of the strong interactions, QCD.
It was not always so. 
As we celebrate the 30 years of the first
publications,
I collect here some memories from those early days. 
I also discuss some of the key issues that were raised back then 
and, despite all progress, are still with us.
\end{abstract}

\vspace{0.5cm}
\begin{center}
{\it Dedicated to the memory of Steven Weinberg, \\
whose ideas will not die, wherever they may lead.}
\end{center}

\end{titlepage}

\section{An Anecdote}
\label{intro}

As I am walking down the hall on the 9th floor
of the RLM building at the University of Texas at Austin
sometime in 1990,
a visibly agitated Steve Weinberg comes out of his office,
makes a beeline for me, and says (approximately) 
``Bira, I just got off the phone with Gerry Brown. 
I called him with some questions because he is one of the statesmen 
in nuclear physics. But all he talks about is fitting data and the rho meson!
It only makes sense to worry about fitting data once you
have a theory that is internally consistent.
You should not just add ingredients to fit data!''
Steve had expressed this point of view many times before
\footnote{For example, ``I do not believe
that scientific progress is always best advanced by keeping an
altogether open mind. It is often necessary to forget one's
doubts and to follow the consequences of one's assumptions
wherever they may lead --- the great thing is not to be free of
theoretical prejudices, but to have the right theoretical
prejudices. And always, the test of any theoretical preconception
is in where it leads.'' \cite{Weinberg:1977book}}.
In fact, this is one of the tenets of a product of Steve's genius,
effective field theories (EFTs) \cite{Weinberg:1978kz}:
you consider the most general dynamics
consistent with the low-energy degrees of freedom and symmetries,
you order these interactions according to the powers of
the ratio of relevant scales that you expect they will contribute to 
{\it observables} (``power counting''),
and you calculate these observables order by order in the expansion.
If you are lucky --- and boy, have we been lucky in nuclear physics! ---
it does {\it not} work as well as you anticipated,
and you learn something.
The context of Steve's phone call was our early struggles 
\cite{Weinberg:1990rz,Weinberg:1991um,Ordonez:1992xp,Weinberg:1992yk}
to formulate
an EFT for nuclear physics, which at the time we referred to
as ``Chiral Lagrangians'' and later as ``Chiral EFT'' because 
of the prominent role played by the light pions and chiral symmetry.
Steve's point was that there should be no particular role for an explicit
rho meson in Chiral EFT: the rho is not a low-energy
degree of freedom since it is unstable with a mass
on the order of the EFT's breakdown scale (and of the masses 
of everything else that is 
strong interacting).

In early 1994, now visiting Austin after having finished my PhD, 
I learn the meaning of the word 
``entranced'' when, out of the blue, I 
get an email from Gerry (via his secretary)
about my proposal with Silas Beane 
for an EFT with quarks and approximate conformal symmetry
\cite{Beane:1994ds}.
Gerry's interest was piqued by a connection to his influential
paper \cite{Brown:1991kk} 
with his close collaborator
Mannque Rho, who had been the first to react positively 
in print \cite{Rho:1990cf} to the nuclear EFT idea.
Unknown to Mannque, Silas and I did most of the work for our 
paper while traveling
to Roma, Lago di Garda, and Trento, where Mannque organized in 
September 1993
one of the first
ECT* workshops, with chiral symmetry as the central topic.
Gerry invites me
for an immediate visit to Stony Brook,
where I find 
an exciting atmosphere, much as in Steve's group. 
For lunch just before my talk,
the whole group congregates 
around a long table in a small library-type room.
After a few bites of food and physics, Gerry suddenly gets up and says 
loudly to me
something like 
``So, young man, you don't know this, but once your advisor
screamed at me on the phone, and now I'm gonna scream at you!''
Before I have a chance to intervene,
Gerry launches onto a long sermon articulating why our lack
of faith on an explicit rho means eternal damnation.
He tops this by
alluding to the unreliability of particle physicists doing nuclear physics
--- echoing a statement in
an earlier paper of his \footnote{
``Occasionally, these days, particle physicists (...) dive into
the meson soup, but come up as soon as they are out of breath, 
brandishing
yet another half-boiled manuscript, running quickly off into
other realms where there is much less data to embarrass them.''
\cite{Brown:1970th}} --- when introducing my talk 
in the same room.
Fortunately I had already planned to start 
with quotes from this \cite{Brown:1970th} and 
another prescient article \cite{Brown:1972ga},
where Gerry pointed out the role of chiral symmetry 
in suppressing two-pion-exchange 
forces and pion-exchange currents.
In the same papers, he urged the implementation of these ideas, which was 
exactly what I was giving a talk about.
I must have made a reasonable case --- or maybe it was just that
I did not give up soon afterwards --- and for the rest of his life
Gerry showed me nothing but support.
And, with time, I understood that Gerry's point
was not only about data fitting --- it also bore on 
the renormalization, and thus on the consistency, of the EFT.

This story is not a fable despite its undertones of Jungian synchronicity.
It presaged much of the cultural rift that has characterized 
the development of nuclear EFTs
in the 30 years that followed:
a tension between effective-field theorists 
searching for a consistent formulation and other nuclear theorists
striving to fit data to ever greater precision,
oftentimes at the expense of EFT tenets.
For the last 20 years there has been an imbalance towards the
second approach \cite{Epelbaum:2008ga,Machleidt:2011zz}, 
in large extent because it is easier to implement
in the new, powerful ``{\it ab initio}'' methods 
for the solution of the Schr\"odinger equation 
(and its few-body variants) \cite{Hergert:2020bxy}. 

I will take here my usual {\it Grillo Parlante} (aka Jiminy Cricket) role
and look back at the early days from the perspective
of the consistency issues that worried us then and are still
around today. These problems have led \cite{Hammer:2019poc}
not only to
a better understanding of power counting in Chiral EFT, but also to
the development of two successful lower-energy EFTs, 
``Pionless EFT'' (where pions are ``integrated out'') 
and ``Halo/Cluster EFT'' (where additionally tight nucleon
clusters are ``integrated in''). 
After recollecting memories of the very early days
(Sec. \ref{notmodel}), 
I remind the reader of some salient implications of EFT to nuclear
physics (Sec. \ref{inter}).
I discuss them in more detail, starting from their roots,
in the following sections:
the justification of power counting 
from renormalization and naturalness (Sec. \ref{renorm});
the importance of few-body forces (Sec. \ref{few-body});
consequences for reactions with external probes (Sec. \ref{probes});
and the role of the Delta isobar (Sec. \ref{Delta}). 
I then conclude (Sec. \ref{conc}).

\section{Glory Days}
\label{notmodel}

In the late '80s and early '90s, Weinberg's Theory Group was a special place, 
and I am not speaking (only) out of nostalgia for the
days of youth. It was not for the mere presence of one of the best
theorists alive. (I have seen theory demigods waiting nervously 
at the door of Zeus's office...)
It was neither just for the company of other superb
theorists --- at various times, Philip Candelas,
Willy Fischler, Vadim Kaplunovsky, and Joe Polchinski --- 
or exceptional graduate students
--- in my year and the next, the likes of Silas Beane, Per Berklund,
Rafael L\'opez-Mobilia, Djordje Minic, and Scott Thomas.
It was that Steve made sure that there was a culture of open discussion
about anything.
Rarely, if ever, one of the weekly Monday brown-bag lunch talks
or Tuesday afternoon seminars would go by without questions
or assessment from Steve.  
Everything one said to the great man was analyzed in a deep, logical
fashion: does it fit with the chain of arguments that starts
with quantum field theory as the consistent combination
of quantum mechanics and relativity?
Everybody, even the string theorists,
spoke EFT, and the ethos was
that any theoretical problem could, should be addressed.

Steve himself was working on a remarkable set of subjects:
string theory, cosmological constant, tests of quantum mechanics,
CP violation, restoration of chiral symmetry, constituent quarks,
and so on.
It came as a surprise to no one, then, when he decided to tackle
nuclear forces from the perspective of EFT.
Steve's first paper \cite{Weinberg:1990rz} laid the ground for the construction
of the nuclear potential and discussed its qualitative properties.
Assuming interactions to have strengths (``low-energy constants'' or LECs)
given by naive dimensional analysis (NDA) \cite{Manohar:1983md,Georgi:1992dw},
scattering amplitudes for processes with at most one nucleon
and momenta comparable to the pion mass, $Q\sim m_\pi$,
are perturbative in a $Q/M_{\rm QCD}$ expansion, 
where $M_{\rm QCD}\sim 1$ GeV
is the characteristic nonperturbative QCD scale \cite{Weinberg:1978kz}.
This sector of Chiral EFT 
is known as Chiral Perturbation Theory (ChPT) \cite{Bernard:1995dp}.
For two or more nucleons, 
Steve identified the possible origin of
the breakdown in perturbation theory that is responsible
for nuclear bound states and resonances:
an $m_N/Q$ enhancement in ``reducible'' diagrams
whose intermediate states contain only nucleons of mass 
$m_N={\cal O}(M_{\rm QCD})$.
He defined the potential as the sum of ``irreducible'' diagrams,
which do not contain this enhancement.
Assuming NDA to apply also to LECs of operators
involving four or more nucleon fields, 
he wrote down the form of the leading two- and three-nucleon
potentials,
leaving 
higher-nucleon forces to the future. 

Steve remembers  \cite{Weinberg:2021exr}
having the insight of applying ChPT ideas to nuclear physics in 1990
while teaching his two-year course on quantum field theory
--- which eventually gave rise to his famous books 
\cite{Weinberg:1995mt,Weinberg:1996kr,Weinberg:2000cr}.
I rarely missed a class and wish I could recall the moment.
I never regarded Steve as a particularly 
dynamic lecturer; what made his lectures
so extremely compelling was the clarity of thought. Each argument
was his own and reconstructed --- for himself as much as the students ---
in real time as he spoke. 
It was common that he paused with a new insight or came back in
the next lecture with a better way of expressing an idea.

What I do recall is that Steve soon gave a brown-bag talk to sharpen
his thoughts about nuclear forces and invite
feedback.
I paid particular attention because for my qualifier
I worked on a little problem
that Steve had suggested earlier:
an estimate of axion production in nucleon collisions.
I was mesmerized by the work of
David Kaplan and others on an EFT for 
axions \cite{Kaplan:1985dv,Georgi:1986df,Georgi:1986kr},
but had to deal also with schematic models for pion production.
It was an ugly match, but sufficient to pass the exam with a happy Steve.
Steve then suggested that I looked
into EFTs for the electroweak symmetry-breaking sector,
with an eye on the treasures promised by the then-viable SSC.
It was soon clear to me that, if I wanted to learn about EFTs
in depth,
I better study them where data already existed to
confront the outcome of power counting.
As Gerry would have expected,
nuclear forces seemed to me just a good warm-up problem.

In his brown-bag talk, Steve echoed a remark from his
paper about the need to go beyond leading
order (LO).
The leading 
potential had some of the ingredients nuclear physicists
had assigned to the two-nucleon interaction --- one-pion exchange (OPE)
and a ``hard core'', represented by one contact interaction in
each $S$ wave.
Mannque's paper \cite{Rho:1990cf}, recommended to me by Steve,
brought up another 
qualitative success of the nuclear EFT idea:
since the leading contact interactions
had no derivatives, exchange currents were expected to be dominated
by pion exchange, in agreement with phenomenology.
Still, the leading 
interactions lacked many features thought to be 
important or even crucial, 
for example
a spin-orbit force, medium-range attraction,
and short-range tensor repulsion.
It was thus not clear whether this approach provided 
a viable foundation for nuclear forces.

Going further looked daunting, and I only jumped in
after a discussion with Carlos Ord\'o\~nez
at the local coffee shop sometime that year.
Carlos, 
then a postdoc wishing to diversify from
string theory, provided the needed reassurance about the importance
of the project.
His research experience, shared cheerfully,
was a crucial guide as we navigated new waters.
At the same time that 
we started working on corrections,
Steve was writing his second nuclear-force paper \cite{Weinberg:1991um},
which gives a fuller account of, and improves on,
the arguments sketched in the first paper.
Its main novelty was a power-counting penalty for the loss of connected pieces,
which suppresses many-body forces 
and makes the LO potential purely two-body.

We quickly found with Steve (Sec. 6
of Ref. \cite{Weinberg:1991um}) that, under NDA,
parity invariance ensures that
next-to-leading-order (NLO) corrections vanish.
This was unfortunate as it meant we had to go to at least
one order higher (N$^2$LO), which
was painful for a couple of reasons.
First, we had started before heavy baryon fields were introduced 
\cite{Jenkins:1990jv}
to simplify the 
counting of derivatives. 
Second, we had to calculate two-pion exchange (TPE) without double counting
iterated OPE. At the time we were using old-fashioned perturbation theory
built from the Hamiltonian, which allows for 
a clean separation of iterated OPE
but requires the evaluation of several noncovariant diagrams for each 
Feynman diagram.
On the up side, the N$^3$LO two-nucleon potential was easy since 
no new contact interactions enter and no iterated OPE needs to be
subtracted.
The isospin-symmetric 
potential we derived \cite{Ordonez:1992xp} 
contained 
the first TPE force consistent with the requirements
of chiral symmetry (no large ``Z diagrams''!).
Already at that point we stressed the role of the 
(Deltaless) N$^3$LO two-nucleon
potential in providing the intermediate-range attraction
usually thought to be relatively strong and due to the sigma meson
and/or correlated pion exchange.
Our potential contained all the ingredients of 
known phenomenological potentials in the form
of seven additional two-nucleon contact interactions,
as well as short-range components in the three-nucleon potential. 

Steve was disappointed that the corrections to the two-nucleon
potential were so complex, and decided in his third paper 
\cite{Weinberg:1992yk}
to emphasize three-body forces, be them
among nucleons only or involving also pions. 
He argued that the leading three-nucleon force actually canceled
against corrections in the two-nucleon force
and focused on pion-deuteron scattering.
He suggested a ``hybrid'' approach where the reaction ``kernel''
--- the external-probe analog to the potential ---
is derived from the chiral Lagrangian and sandwiched
between wavefunctions obtained from a (mostly two-nucleon) 
potential taken from elsewhere.
After another of Steve's brown-bag talks in the spring of 1992,
Ricardo Mastroleo, 
a visiting professor from Brazil,
and I helped Steve with the evaluation of 
pion-deuteron integrals. 

Now, contrary to Steve, I thought the N$^{3}$LO potential
was promising. In addition to being accompanied 
by a prescribed three-nucleon force,
the two-nucleon potential was not more complicated
than most ``realistic'' phenomenological potentials.
I expected that this would be sufficient to 
spur further work and I could move on to other things.
Fortunately at UT there was an extraordinary nuclear theorist, Lanny Ray,
who convinced me that if I wanted to convert others to the gospel,
I had to walk the line and fit two-nucleon data.
I got similar encouragement 
later from, among others:
Dave Ernst, when Carlos and I visited Texas A\&M University circa 1992; 
Johan de Swart, Thom Rijken, Rob Timmermans, and the rest of the Nijmegen group,
when they welcomed me as family in the summer of '92;
and Karl Holinde, when he visited Seattle
a couple of years before his untimely passing.
To our N$^3$LO potential from Ref. \cite{Ordonez:1992xp},
Carlos and I added TPE with intermediate Delta isobars,
and 
Lanny adapted the methods he had developed for 
a phenomenological two-nucleon potential with Delta and Roper \cite{Ray:1987ir}.
In my PhD thesis 
\cite{VanKolck:1993ee},
I reported preliminary results of a fit to the
SAID phase shifts (based on Ref. \cite{Arndt:1986jb})
at one cutoff.

It became obvious when I ventured away from Austin that
Lanny had a fine perception of the nuclear community:
it took good notice of the approach but was, with a few notable exceptions,
skeptical that it would succeed. 
Starting with my postdoc interview talks,
I was told in no uncertain terms by many people (besides Gerry) 
that our replacement of heavier-meson exchange by contact interactions
had no chance to work, even for the deuteron!
I was fortunate to find shelter in the Nuclear Theory Group
at the University of Washington.
Under Jerry Miller's leadership,
the group fostered new ideas and encouraged its junior members to
work independently if they chose to 
--- an environment where I could polish, enlarge,
and publish parts of my thesis, as well as tackle new challenges.
Despite his boundless creativity and stamina, Jerry knew 
how to strike the perfect balance between stimulation and 
expectation. 

The attitude in the community 
started to change only after we published the fit to 
two-nucleon phase shifts and deuteron properties.
The first publication \cite{Ordonez:1993tn} was essentially
the corresponding chapter in my thesis. 
The second paper \cite{Ordonez:1995rz} contained the full 
details about the potential and the solution of 
the Schr\"odinger equation.
We also presented fits to the then-new
Nijmegen partial-wave analysis \cite{Stoks:1993tb}
for three different cutoff values.
As we emphasized, we had no hope of matching fitting standards
of the day, which called for a $\chi^2/{\rm dof}\simeq 1$ up to
the pion production threshold.  
The ``realistic'' potentials that achieved this 
were teleological: they added 
the ingredients needed to reach this objective.
In contrast, our point was that 
a decent fit follows from the theoretical prejudices of EFT.
Simplicity was not in the form of the cake, but in its recipe.

One of the first to realize the promise of this approach was Jim Friar,
who visited Austin in 1991 or 1992 and immediately 
became a vocal supporter. In fact, he had anticipated many
of the EFT power-counting arguments \cite{Friar:1977xh,Coon:1986kq}. 
During his visit he taught us much about nuclear issues not usually stated 
explicitly in papers.
Jim's passion for physics meant that he 
would walk the line with me.
I had not met a senior theorist with Jim's dedication to, and ability
for, long analytical calculations. I frequently found myself
trying to catch up to him.
With various collaborators in the following years,
we elaborated on two 
topics that I had only sketched in my thesis: few-nucleon forces
and isospin violation.

Jim explained to us that it was well known --- at least to those who knew it
well... --- that energy dependence in the two-body potential affected
three-body forces. In Chiral EFT, this translates
into a cancellation of the corrections to the LO two-nucleon
potential that are linear in the energy against both
the leading three-nucleon and double-pair diagrams. 
The cancellation only went through if Steve's three-body force
contained a small error. Steve's reaction when I told him 
was characteristic
of his supreme intellectual honesty: he just accepted it with
a modest comment. In fact,
despite the enormous achievement gap between us 
he never ever used an argument based on authority,
which of course contributed a great deal to my growth as a theorist.
The consequence of the cancellation is that the dominant three-nucleon
force would come at N$^2$LO from the Delta, followed by 
the N$^3$LO force from interactions in the NLO Lagrangian
\cite{vanKolck:1994yi}.
Among those, TPE is determined by chiral
symmetry with a form that had not been properly accounted for before
\cite{Friar:1998zt}.
The shorter-range components, in turn, were shown not to be
negligible \cite{Huber:1999bi}.

Jim was also very interested in isospin violation.
In my thesis I emphasized that isospin violation is an
accidental symmetry \cite{vanKolck:1995cb}, 
namely an approximate symmetry of
the EFT at LO that it is not a symmetry
of the underlying theory, here QCD.
In my first paper with Jim and Terry Goldman \cite{vanKolck:1996rm}, 
we essentially classified isospin-violating
nuclear forces and discussed the most salient interactions.
For many years afterwards, we 
derived various components
of the isospin-breaking two- and three-nucleon potentials
\cite{vanKolck:1997fu,Friar:1999zr,Friar:2003yv,Friar:2004ca,Friar:2004rg},
which complement the isospin-symmetric potentials of 
Refs. \cite{Ordonez:1992xp,Ordonez:1993tn,Ordonez:1995rz}.

A final aspect of the early days were reactions.
One late Friday afternoon 
while shooting the breeze with other UT graduate students 
I had the epiphany that EFT provides the basis for not only nuclear forces
but the whole ``traditional'' nuclear physics.
Silas joined enthusiastically.
He reminds me that I gave him a draft of Steve's third paper,
where he found a factor-of-2 error that Steve had apparently
already caught. 
As I heard from Jim and Justus Koch,
who opened the doors of NIKHEF and his home to me in the summer of '92, 
there was
enormous interest in the 
theoretical interpretation of the 
new threshold data from Mainz and Saskatoon on
neutral-pion photoproduction at threshold --- in particular,
the novel predictions from ChPT at loop level 
\cite{Bernard:1995dp} in disagreement
with tree-level phenomenology.
For the same reason that loops are relatively important in the 
single-nucleon sector --- the vanishing of LO --- 
threshold neutral-pion photoproduction on the deuteron
should be sensitive, Silas and I reasoned, to two-nucleon effects.
Silas enlisted a younger UT student, Chang-Yong Lee,
in an N$^2$LO calculation.
Despite the encouragement from Mannque, Ulf Mei{\ss}ner, and especially the
late, larger-than-life Aron Bernstein,
the project evolved slowly and was published 
\cite{Beane:1995cb} only after Silas and I had left Austin.
Still, it served as the basis to go one order further a couple of years
later \cite{Beane:1997iv}.
At this order the proton amplitude agreed well with data,
and Silas's insight was that the ChPT amplitude for the neutron
could be tested indirectly. 
An early success of nuclear EFT was the confirmation
of our prediction (within error bars from
the EFT truncation) by a measurement at Saskatoon 
soon afterwards \cite{Bergstrom:1998zz}.
Encouraged by this success we carried out later with Manuel Malheiro,
Judith McGovern and Daniel Phillips
a program for the extraction of
neutron polarizabilities from Compton scattering on the deuteron
\cite{Beane:1999uq,Beane:2002wn,Beane:2004ra},
which has eventually met with similar success 
\cite{Griesshammer:2012we}.
(See also Daniel's contribution to this volume \cite{Phillips:2021yet}.)

In the meantime, Mannque had assembled a team
with his Seoul collaborator Dong-Pil Min and 
his exceptional student, Tae-Sun Park.
Perhaps influenced by Steve's assessment about the subleading potential, 
they carried out hybrid computations 
of axial \cite{Park:1993jf,Park:1994ai} and 
vector \cite{Park:1994sr,Park:1995pn}
currents, applying the latter to radiative
neutron-proton capture.
Tae-Sun does not get the recognition he deserves for this pioneering 
work, perhaps because he was partly out of sight.
Mannque did what he could to promote both Tae-Sun's and my work,
inviting me not only to speak at the 1993 ECT* workshop on chiral symmetry
but also to lecture at a summer school in Seoul in June 1994.
The warm Korean hospitality helped me keep going as 
a young postdoc.

In order to convince skeptics, we were always on the lookout for opportunities
to go beyond conventional models and, 
as I was finishing up my PhD, 
Lanny drew my attention to problems in the
theoretical explanation for new data on
neutral-pion production in nucleon-nucleon collisions near threshold.
This was closely related to my qualifier problem, and
Seattle was the right place to land on:
Jerry had done the benchmark calculation for
this process \cite{Miller:1991ndb}.
With Jim and Tom Cohen, who happened to be visiting on sabbatical,
we immediately realized \cite{Cohen:1995cc}
that the power counting had to be modified on
account of the larger typical momentum ${\cal O}(\sqrt{m_\pi m_N})$
required to create a pion from a two-nucleon state.
As a consequence of a vanishing LO and other cancellations,
the bulk of the
cross-section arises from contributions that are relatively 
unknown, which we modeled by heavier-meson exchange
\cite{vanKolck:1996dp}.
With a visitor from Brazil, Carlos da Rocha,
Jerry and I marched on to charged-pion production, which is less pathological 
but still hard to describe \cite{daRocha:1999dm}.
As then-postdoc Chris Hanhart showed, $P$ waves are better behaved 
and can be used to constrain the three-nucleon force
\cite{Hanhart:2000gp}.
In parallel, Kuniharu Kubodera and Fred Myhrer,
with whom I had discussed EFT after they drove to Duke University 
during my 1994 visit, 
analyzed neutral-pion production 
with the standard power counting designed for momenta ${\cal O} (m_\pi)$
\cite{Park:1995ku}. 
We eventually  
collaborated in order to understand the peculiarities of 
this reaction \cite{Hanhart:2000wf}.
The biggest success was tracking the effects
of charge-symmetry violation from the quark mass difference.
We enlisted Jouni Niskanen for a prediction of the forward-backward
asymmetry in $pn\to d\pi^0$ \cite{vanKolck:2000ip}, which
was confirmed experimentally within a couple of years at TRIUMF 
\cite{Opper:2003sb}.
We then extended this work to the more complicated $dd\to \alpha \pi^0$,
which was detected for the first time at the level we expected 
\cite{Stephenson:2003dv}.
We later improved the treatment of the various contributions
to this complicated process \cite{Gardestig:2004hs,Nogga:2006cp}.

Within a few years of the first publications, 
the case had been made, it seemed to me \cite{vanKolck:1999mw}, for 
nuclear EFT as a foundation for the whole low-energy nuclear
physics:
many qualitative aspects
of traditional nuclear physics had been explained, 
a decent (but not spectacular) fit to two-nucleon data obtained,
useful currents derived,
and predictions confirmed experimentally.
Yet, the fun was just beginning.

\section{The Age of the Cockroaches}
\label{inter}

Until about 1996, nuclear EFT felt like a lonely place.
I believe I have already mentioned everybody worldwide who had an
active interest in the subject, and most of them were focused on specific 
processes rather than a comprehensive approach.
Then, catalyzed by the challenges raised in 
Refs. \cite{Kaplan:1996xu,Phillips:1996ae,Phillips:1997xu,Beane:1997pk}
and confronted in two workshops (1998 in Pasadena \cite{Seki:1998qw} 
and 1999 in Seattle \cite{Bedaque:2000kn}),
a small nuclear EFT community started to form.
There are plenty of anecdotes from this era, but I save them 
for when we celebrate it.

1999 witnessed the clash in an epic Trento workshop (co-organized with
Rob, Paulo Bedaque, and Ben Gibson) between
young effective-field theorists and established potential modelers,
who traded the war names of ``cockroaches'' and ``dinosaurs''.
Trento was the place where many in the nuclear theory community 
had the first opportunity to see EFT in depth.
EFT was increasingly becoming part of the mainstream.
The first signal was Jim's work with the Nijmegen group
\cite{Rentmeester:1999vw,Rentmeester:2003mf},
who successfully replaced the one-meson-exchange interaction in their
state-of-the-art phase-shift analysis \cite{Stoks:1993tb}
by the long-range part of chiral TPE.
The process picked up pace 
with the improvements made in the Deltaless potential by 
the Bochum-J\"ulich group and their collaborators
\cite{Epelbaum:1999dj,Epelbaum:2002vt},
and then accelerated dramatically after the landmark
fit to two-nucleon data by David Entem and Rupert Machleidt \cite{Entem:2003ft}.
These developments 
came at a time when {\it ab initio} many-body methods were mature enough
to welcome, and soon test, two- and three-nucleon forces
having connections with QCD. 
Nowadays, rare is the {\it ab initio} calculation
that does not use a chiral potential as input.
We have reached the point where predictions for a wide variety of nuclei 
can be confronted with data and used to infer
the pros and cons of the input potentials \cite{Hergert:2020bxy}.

As great as this advance is, it has been 
patterned on the early work, with nuclear potentials and currents
derived from EFT to some order but then treated as black boxes,
as if they had originated in just another model. 
Since 1996, however, 
we have been discovering the various ways in which
this approach is at odds with the tenets of EFT:
\begin{itemize}
\item 
In EFT, there is a hierarchy among the components of the potential.
LO should capture the essential physics, for example
generating bound states and saturation
at momenta (or densities) within the regime
of the theory.
Subleading orders, on the other hand, should be amenable to perturbation 
theory. Because EFT potentials are increasingly singular with order,
it is not guaranteed that it makes sense
to solve the Schr\"odinger equation exactly
with a potential truncated at some subleading order.
I discuss the insistence on iterating what does not need to, and probably 
cannot, be iterated --- which I like to call the nuclear Cohen syndrome
in homage to Tom, who identified the disease ---
in Sec. \ref{renorm}.
\item
The attitude ingrained in the nuclear community is that one needs to nail
the two-nucleon potential down perfectly up to some 
high energy before considering three-body forces, the form of which
is difficult to guess phenomenologically. 
Aptly called $\chi^2$ paranoia by Rupert,
this attitude persists today in the effort to go to very high order in the
chiral two-nucleon potential --- see Rupert's 
perspective in this celebratory volume \cite{Machleidt:2021ggx}.
There is of course nothing wrong with trying to describe the
two-nucleon system better, but one should keep in mind that the 
EFT expansion is in a ratio of momentum scales, 
not in the number of nucleons {\it per se}.
It might be possible to get a good description of nuclei 
including the few-body forces that are prescribed by the theory
without attaining at the same order a perfect fit to two-nucleon
scattering all the way to the pion-production threshold. 
This is the topic of Sec. \ref{few-body}.
\item
EFT is a theory of everything (in low-energy nuclear physics): 
the same remarks about the potential apply
to the kernels of reactions involving external probes
like pions and photons, see Sec. \ref{probes}.
\item
The fact that pions exist
does not mean that they need to appear explicitly in the theory.
But if they do, then another low-energy degree of freedom 
--- the Delta isobar --- should also be important. 
Delta denialism is confronted in Sec. \ref{Delta}.
\end{itemize}

\section{Renormalization and Naturalness}
\label{renorm}

Steve's papers do not betray much worry about renormalization,
but that was certainly in the air in Austin.
At some point while walking through campus 
I had a panic attack when it downed on me 
that the OPE tensor force might lead to extreme
regulator dependence in the solution of the Schr\"odinger equation.
I calmed down when I
convinced myself that just looking at the diagrams
in perturbation theory was not a good guide for what happened
with the full, nonperturbative solution.
In his second paper \cite{Weinberg:1991um}, 
Steve solved analytically the case of a pure,
non-derivative contact interaction, 
from which it was clear that renormalization
was different from that of the perturbative series.
The problem was renormalizable
despite the singularity of the Dirac delta function,
giving us hope that it would be so 
also with the additional OPE at LO.
If that was the case, then we expected 
a truncation of the potential at some order to yield all terms
of the truncation of the amplitude at the same order, plus
a subset of small, harmless higher-order contributions.

In addition to the finite number of irreducible loops in a truncation
of the potential, the amplitude contains also
the reducible loops subsumed in the
solution of the Lippmann-Schwinger
equation --- or, equivalently, the Schr\"odinger equation. 
A loop requires a regulator to suppresses high momenta.
The favorite technique is dimensional regularization,
but it can only be implemented when the calculation
is reduced analytically to a finite number of loops
--- for example, the contact interaction Steve solved.
In an exact numerical solution of the Schr\"odinger equation, as needed
for LO OPE, one is effectively summing up an infinite number of loops.
A loop is a loop, so it is natural to use 
for the potential and the dynamical equation
the same cutoff regulator: loop integrals are defined 
with a function $f(p/\Lambda)$ of momentum $p$ 
parametrized by a single number
$\Lambda$, such that $f(0)=1$ but $f(\infty)=0$.
Back then \cite{Ordonez:1993tn,Ordonez:1995rz}, 
we chose a Gaussian function of relative momentum 
to make the potential as local as we could
while killing
all potential divergences.
This regulator also allowed us to check
parts of the long-range potential against
Rijken's papers \cite{Rijken:1990qs,Rijken:1991jm}.
Soon, the potential was recalculated in dimensional regularization 
\cite{Kaiser:1997mw,Kaiser:1998wa}, which is sufficient for
perturbation theory in high partial waves
where the centrifugal barrier prevents two low-energy
nucleons from getting close. 
The long-range part of the potential 
\cite{Rentmeester:1999vw,Rentmeester:2003mf}
is the same, but its 
overall form much simpler.
Most subsequent work adopted different regularizations for 
irreducible and reducible loops, with
cleaner expressions but dirtier regulator dependence.

The model independence of EFT --- hence its claim to reproduce QCD at 
low energies --- is only guaranteed for
{\it observables}, 
and only if the effects of the arbitrary regulator are removed by 
renormalization. How can they be removed? In pre-history, it was 
thought necessary to send the cutoff $\Lambda$ to infinity, 
but in EFT it is sufficient that its effects
be suppressed by positive powers of $Q/\Lambda$ 
for $Q\sim M_{\rm lo}\ll M_{\rm hi}$,
where $M_{\rm lo}$ stands for the low-energy scales
and $M_{\rm hi}$ for the breakdown scale.
The aim of an EFT is to construct the most general $S$ matrix,
with a power counting that allows for controlled truncations at each order $n$:
\begin{equation}
S^{(n)}(Q\sim M_{\rm lo}) = 1 + \sum_{\nu= 0}^{n} \, 
 \left(\frac{Q}{\Mhi}\right)^{\!\nu} \, 
 F^{(\nu)}\!\left(\frac{Q}{\Mlo}; \gamma^{(\le \nu)}
 \right) 
 + {\cal O} \left(\frac{Q^{n+1}}{M_{\rm hi}^{n+1}}, 
\frac{Q^{n+1}}{M_{\rm hi}^{n}\Lambda}\right)
\,,
\label{Texp}
\end{equation}
where the $F^{(\nu)}$s are functions generated at order $\nu$ 
by the dynamics of the explicit degrees of freedom
and $\gamma^{(\le \nu)}$ stands for the LECs appearing up to that order.
Then, for sufficiently large $\Lambda$
--- not necessarily only for $\Lambda \to \infty$ \cite{Lepage:1989hf} ---
regulator effects can be made no larger
than the next order in the EFT expansion, which is of relative
size ${\cal O}(Q/M_{\rm hi})$. 
When LO is nonperturbative often we have to check renormalizability numerically,
which might require values well above $M_{\rm hi}$ to rule out
mild dependence, such as logarithmic. 

In order to achieve $\Lambda$-independent $F^{(\nu)}$s in Eq. \eqref{Texp}, 
enough LECs need to be present at each order
to remove the non-negative powers of $\Lambda$ generated by the loops.
Naturalness \cite{tHooft:1979rat,Veltman:1980mj} 
is the idea that a LEC needed to absorb the 
cutoff dependence reflects sensitivity to short-range physics
at the same order. The LEC then should not only remove
the offending powers of $\Lambda$ from observables but also have 
an unknown finite part. The latter is
to be determined by matching to either the underlying theory or to
data.
The NDA \cite{Manohar:1983md,Georgi:1992dw}
on which Weinberg's power counting is based was derived by looking 
at scattering amplitudes perturbatively, renormalizing them with the necessary
LECs, and assuming naturalness.
Steve applied this idea to the potential, which he argued 
has a perturbative expansion.
The hope in early days was that the nonperturbative part of the calculation
would not generate further non-negative powers of $\Lambda$
and thus not affect the NDA-based ordering of interactions.

With our local regulator \cite{Ordonez:1993tn,Ordonez:1995rz}, 
every contact interaction
contributed to every wave.
To facilitate the comparison with one-boson exchange models, 
we did not use in the fit the minimum 
number of nine two-nucleon LECs that appeared up to N$^{3}$LO 
according to NDA \cite{Ordonez:1992xp}, 
but a redundant set of 18.
Starting with Ref. \cite{Epelbaum:1999dj}, the minimal set 
of LECs and nonlocal regulators for the dynamical equation 
have been frequently employed, 
allowing for much easier fits partial wave by partial wave.
But, all in all, in 1996 
we had 26 parameters and the fitting procedure needed at each
cutoff was laborious. 
The qualitatively similar results for the Deltaful N$^3$LO potential
at the three cutoffs of $\simeq$ 500, 780 and 1000 MeV 
we could reasonably handle
\cite{Ordonez:1995rz} gave us some reassurance
that the problem was renormalizable. 
No panic attack.

It did not take long, though, for 
David, with Martin Savage and Mark Wise,
to point out that there is a problem:
in the two-nucleon $^1S_0$ channel, one can see semi-analytically
\cite{Kaplan:1996xu,Beane:2001bc,Long:2012ve} 
that at LO in the original power counting an 
$m_\pi^2\ln \Lambda$ dependence
is generated by the dynamical equation which cannot be absorbed
in the chiral-symmetry-conserving LEC present at that order.
The four-nucleon-field interaction that removes this cutoff dependence 
breaks chiral
symmetry explicitly in the same way as the average quark mass in QCD,
that is, as a component of a chiral four-vector.
As a consequence, it is 
accompanied, unlike the chiral-symmetric interaction
present at LO, by an infinite tower of additional
interactions involving an even number of pions. 
It should only appear at N$^2$LO according to NDA,
so clearly NDA --- the cornerstone of the approach --- is being violated. 

For a while the wishful thinking was that this could be an isolated failure,
or at most limited to chiral-symmetry breaking.
If so, then some contributions would be enhanced, but generally
only in higher orders --- for example, in two-nucleon scattering 
when the additional pions formed
loops --- and/or in less fundamental processes --- like pion scattering
on the three-nucleon system. 
But then came the work of Daniel, Tom, and Silas
\cite{Phillips:1996ae,Phillips:1997xu,Beane:1997pk},
which showed that, when pion interactions are turned off,
Weinberg's prescription does not work at subleading orders.
The importance of crises in scientific progress is
well documented, and a crisis this was.

The community has been split ever since. 
Most people decided to 
simply choose regulators (spectral-function regularization,
``super Gaussian'' functions, different cutoff values for each
partial wave, {\it etc.}) that allow for the best data fitting,
along the way giving up on the idea of {\it a priori} error estimates.
This attitude has produced numerous phenomenological successes
\cite{Epelbaum:2008ga,Machleidt:2011zz} 
as well as some failures which are gaining increased attention.
Effective-field theorists have, in contrast,
fussed over the issue of renormalization.

It became clear soon that the nuclear Cohen syndrome
can be traced to treating the subleading parts of the potential
nonperturbatively 
\cite{vanKolck:1997ut,Kaplan:1998tg,Kaplan:1998hb,vanKolck:1998ra,Kaplan:1998we,vanKolck:1998bw}. 
LO does require an exact solution to produce
the $S$-matrix poles we associate with bound states and resonances,
but subleading interactions should in principle be incorporated 
in distorted-wave perturbation theory.
In the case of purely contact interactions, 
naturalness at LO implies that the poles are outside the range of the theory.
Once the fine tuning required for the shallow
deuteron and $^1S_0$ virtual state
is incorporated, naturalness 
still allows for the 
construction of a consistent power counting for subleading interactions.
This Pionless EFT \cite{Hammer:2019poc} works well 
at momenta below the pion mass. 

The first important lesson from Pionless EFT is thus that, while one
can expand the potential,
each term enters the expansion of the amplitude differently.
A FAQ is why, if a correction is, well, a correction, can it not be treated
nonperturbatively?
The answer is that in general only for regular interactions can one do
this while preserving model independence.
For the increasingly singular EFT interactions,
what is small 
is the sum of contributions at a given subleading order in the amplitude. 
There can be cancellations, the magnitude 
of individual contributions being regulator dependent. 
When one solves exactly for a subleading truncation of the potential, 
one includes in the amplitude 
many loops without the LECs that would cancel their 
cutoff dependence. 
Sometimes --- for example, implicitly when using as input 
to LO Chiral EFT data at physical pion mass 
rather of their chiral-limit values --- 
this can be done without affecting
renormalization.
However, resumming (``unitarizing'') a truncated EFT potential 
will usually generate regulator dependence: 
the partial subset of higher-order corrections is not harmless.
Unless resummation with a specific regulator can be somehow justified,
the resulting regulator dependence must be considered 
a model for the short-range physics.
The question that should be frequently asked instead is, 
if corrections are indeed corrections, should they not 
be amenable to distorted-wave
perturbation theory, regardless of regularization issues?
There are no tests that this is the case for NDA-based chiral potentials.

The second important lesson for Chiral EFT is the power counting
of factors of $4\pi$ in reducible loops.
In addition to the $m_N/Q$ enhancement noticed by Weinberg,
there is a second enhancement of $4\pi$. For example, one should count
\begin{equation}
\text{reducible two-nucleon loop} \sim \frac{m_N Q}{4\pi} \,,
\end{equation}
instead of the $Q^2/(4\pi)^2$ for an irreducible loop.
In Pionless EFT the LO contact interaction
is fine tuned to be (after renormalization, of course)
\begin{equation}
\text{nonderivative contact interaction} \sim \frac{4\pi}{m_N M_{\rm lo}},
\end{equation}
so the LO loops in the Lippmann-Schwinger equation
are an expansion in
$Q/M_{\rm lo}$.
The dynamical equation needs to be solved nonperturbatively
and can lead to poles at momenta $\kappa_A={\cal O}(M_{\rm lo})$.
OPE, on the other hand, has a strength 
\begin{equation}
\text{one-pion exchange} \sim \frac{4\pi}{m_N M_{N\!N}},
\end{equation}
where $M_{N\!N}\equiv 16\pi f_\pi^2/g_A^2 m_N$ 
\cite{Kaplan:1998tg,Savage:1998vh,Kaplan:1998we}, 
with $f_\pi\simeq 93$ MeV the pion decay constant 
and $g_A\simeq 1.27$
the LO pion-nucleon coupling. The OPE ladder is 
thus a series in $Q/M_{N\!N}$.

David, Martin, and Mark 
built on Pionless EFT
by suggesting that $M_{N\!N}$, which numerically is $\approx 300$ MeV,
be treated as a high-energy scale 
\cite{Kaplan:1998tg,Savage:1998vh,Kaplan:1998we}.
In this case LO is formally the same as in Pionless EFT,
pions can be treated
perturbatively, and no renormalization issues arise from
pion exchange. Unfortunately, convergence of perturbative
pions deteriorates rapidly for momenta above 100 MeV or so 
\cite{Fleming:1999ee,Kaplan:2019znu},
and at physical pion mass this version of Chiral EFT does not greatly
improve over Pionless EFT.

In hindsight, such a failure is not unexpected:
from naturalness, $m_N={\cal O}(M_{\rm QCD})$,
$f_\pi={\cal O}(M_{\rm QCD}/4\pi)$,
and $g_A={\cal O}(1)$,
so that 
$M_{N\!N}= {\cal O}(f_\pi)$.
The two reducible-loop enhancements 
lead to the expectation that OPE needs to be resummed and generates
poles with momenta
\begin{equation}
\kappa_A = {\cal O}(f_\pi) ={\cal O}\left(\frac{M_{\rm QCD}}{4\pi}\right) \,,
\label{binding}
\end{equation} 
in the absence of fine tuning.
While this is somewhat larger than the binding momentum of the deuteron
($\kappa_2\simeq 45$ MeV), it is not too far off for heavier nuclei.
Of course this argument is very simplistic as it does not take into account
the range of the potential given by the pion mass.
But, as Steve pointed out \cite{Weinberg:1991um}, 
without the additional enhancement by $4\pi$ one would 
have to consider $m_N$ as particularly large,
in fact ${\cal O}(M_{\rm QCD}^2/M_{\rm lo})$ \cite{Ordonez:1995rz}.
This dated argument underlies most subsequent work with chiral potentials
\cite{Epelbaum:2008ga,Machleidt:2011zz},
where nucleon recoil and relativistic corrections are taken to
be suppressed by powers of $M_{\rm QCD}^2$
and the origin of $M_{\rm lo}$ is left in the air.
In contrast, Eq. \eqref{binding} suggests 
$M_{\rm lo}={\cal O}(f_\pi)$
and explains why $\kappa_A \ll M_{\rm QCD}$, a mystery when one does
not take into account naturalness in the presence of
spontaneous symmetry breaking.

But can we renormalize the tensor part of OPE, which does not contribute to 
$^1S_0$? 
Memories of my Austin panic attack haunted me when Martin argued in talks
--- for example, Ref. \cite{Savage:1998vh} ---
that successive loops in the OPE ladder would require 
LECs from interactions with more and more derivatives.
So it was high time to settle this issue in the nonperturbative context.
It is remarkable that the LO in Weinberg's power counting 
{\it is} renormalizable
in the coupled $^3S_1$-$^3D_1$ channels after all 
\cite{Frederico:1999ps,Beane:2001bc,Eiras:2001hu,PavonValderrama:2005gu}.
Unfortunately, this is not true
in the higher waves where the OPE tensor force is attractive  
\cite{Nogga:2005hy,PavonValderrama:2005uj}
but Weinberg's power counting prescribes no LO LECs.
The solution is to treat pions at LO only in the lower waves 
--- besides $S$ waves, $^3P_0$ and perhaps a couple more --- 
where it
is nonperturbative, together with the LECs that guarantee renormalization 
\cite{Nogga:2005hy}.
OPE in higher waves is suppressed by powers
of the orbital angular momentum due to the centrifugal barrier 
\cite{Birse:2005um,PavonValderrama:2016lqn,Kaplan:2019znu}
and can be included in higher orders.
More singular corrections can indeed 
be renormalized in distorted-wave
perturbation theory \cite{Birse:2007sx,Long:2007vp,Birse:2010jr}.

Reasonable agreement with data is found
\cite{Valderrama:2009ei,Valderrama:2011mv,Long:2011qx,Long:2011xw,Long:2012ve,Wu:2018lai},
but clearly improvements are needed, especially in the $^1S_0$ channel
\cite{Beane:2001bc,Long:2013cya,SanchezSanchez:2017tws}.
Fixing the chiral expansion opens up the possibility of matching to
lattice QCD data at pion masses where Chiral EFT converges.
A possible scenario \cite{Beane:2001bc} is that the deuteron
and $^1S_0$ virtual state accidentally cross zero energy 
close to physical pion mass, a description of fine tuning 
in terms of quark masses 
analogous to that of atomic Feshbach resonances in terms of an 
external magnetic field.

But these ideas also have consequences beyond the two-nucleon system...

\section{Few-Body Forces}
\label{few-body}

In the first paper \cite{Weinberg:1990rz}
Steve did not realize that there is an
additional suppression for few-nucleon forces with respect to two-nucleon
forces.
He was quick to rectify this in his second paper \cite{Weinberg:1991um},
where he argued for a $Q^2/M_{\rm QCD}^2$ suppression for 
each additional nucleon involved in the potential.
This argument was based on a particular counting of factors of $4\pi$.
Already in 1996, Jim proposed \cite{Friar:1996zw} a different counting 
where the penalty is instead $Q/M_{\rm QCD}$.
Jim based his argument on his pre-EFT ideas \cite{Friar:1977xh,Coon:1986kq}, 
but it is also what results when one counts $4\pi$s
in reducible loops as in Pionless EFT \cite{Hammer:2019poc}.
In either case, the relative suppression
of few-nucleon forces was an early qualitative success of Chiral EFT,
as it matched experience with phenomenological models that include 
pion exchange.

However, the relative importance
of few-nucleon forces is not well established in Chiral EFT.
It is worrisome that saturation in nuclear matter with 
NDA-based chiral potentials
seems to occur in the low-energy region only at orders 
where three-body forces are included 
\cite{Drischler:2017wtt,Sammarruca:2018bqh}. 
A careful uncertainty quantification \cite{Drischler:2020yad}
indicates that NDA-based chiral potentials have such large
errors at LO and NLO that any equilibrium density is possible.
There is no guarantee that the equilibrium point of symmetric nuclear matter
is within Chiral EFT but, if it is, 
it must appear in its region of applicability
already at LO, to be moved only slightly by higher orders. 
(Nobody would --- or at least should ---
buy into an expansion where the alpha particle, say,
does not have at LO
a definite binding momentum below the breakdown scale.)

In fact, in Pionless EFT \cite{Hammer:2019poc} a three-body force is required 
at LO to achieve renormalization in the three-body system 
\cite{Bedaque:1998kg,Bedaque:1998km,Bedaque:1999ve}, which 
then explains Efimov physics \cite{Braaten:2004rn}.
It is sufficient for renormalization in larger systems also, without
the need for four-body forces \cite{Platter:2004he,Platter:2004zs}
until one reaches NLO \cite{Bazak:2018qnu}. A good description of
systems with $A=3,4$ nucleons emerges starting with
two-body unitarity at LO \cite{Konig:2015aka,Konig:2016utl,Konig:2019xxk}.

This example of proper nuclear structure without
a fit to high-energy two-nucleon data
has not been sufficient to cure $\chi^2$ paranoia.
But it has led Alejandro Kievsky and collaborators to propose
that three-nucleon forces should be LO also in Chiral EFT 
\cite{Kievsky:2016kzb,Kievsky:2018xsl}.
They have pointed out that the description of three-nucleon data 
with NDA-based chiral potentials is 
relatively poor even at the high orders 
where the two-nucleon system is essentially perfectly fitted.
Adding a three-body force improves on Weinberg's LO 
significantly for $A=3,4$ \cite{Kievsky:2016kzb}
and yields nuclear saturation \cite{Kievsky:2018xsl}. 
The difference with Pionless EFT is that 
there is no evidence the three-nucleon force is needed for LO renormalization
once the two-nucleon system is renormalized 
\cite{Nogga:2005hy,Song:2016ale,Yang:2020pgi}.
In this case, results are 
reasonable for $A=3,4$ up to NLO without three-body forces 
\cite{Song:2016ale,Yang:2020pgi},
but at LO $p$-shell nuclei are unstable \cite{Yang:2020pgi}
and nuclear matter underbound \cite{Machleidt:2009bh}
for the limited cutoff values that are accessible.
Jerry Yang and collaborators have suggested recently 
\cite{Yang:2021XXX}
that three- and four-nucleon
forces might be enhanced by combinatorial factors of $A$.
I do not think we have heard the last of enhanced few-nucleon forces.

\section{External Probes}
\label{probes}

Concern existed for a long time
about the consistency between reaction kernels and potentials. 
It was hoped that Weinberg's hybrid approach 
would capture the most important physics, but 
we were aware that it would have to be replaced
as higher precision was sought. 
Already in our first reaction paper \cite{Beane:1995cb},
we used both chiral \cite{Ordonez:1993tn}
and Bonn \cite{Machleidt:1987hj} wavefunctions
in an attempt to gauge the uncertainties caused by the mismatch
between potential and kernel. 

One issue is that observables must be independent of the choice
of fields, which means they should be obtained consistently
from the same Lagrangian.
Model independence also requires proper renormalization, 
just as for
scattering amplitudes involving no external probes.
One can expand the long-range part of a kernel with Weinberg-type counting,
but the sizes of short-range interactions are not necessarily given
by NDA. 
Power counting is for the amplitude, and the LECs of short-range interactions
involving external probes 
might also be enhanced by renormalization running.
That this is indeed the case was shown by 
Daniel and 
Manolo Pav\'on Valderrama \cite{Valderrama:2014vra}. 

A concrete example is neutrinoless double-beta decay,
a lepton-number-violating process
where two neutrons change into two protons and two electrons.
It is expected to be driven by the exchange of a Majorana neutrino,
whose tiny mass is generated by the lowest-dimensional (dimension-five) 
operator in the Standard Model EFT \cite{Weinberg:1979sa}. 
The most important contribution comes from 
the $^1S_0$ channel, where its renormalization is similar to
OPE's \cite{Kaplan:1996xu,Beane:2001bc,Long:2012ve}.
The consequence is that a short-range lepton-number-violating 
interaction is needed at LO \cite{Cirigliano:2018hja,Cirigliano:2019vdj},
but no new interaction appears at NLO \cite{Cirigliano:2019vdj}. 
The corresponding LEC can be estimated from the isospin-violating 
electromagnetic interactions to which it is related by chiral symmetry
\cite{Cirigliano:2018hja,Cirigliano:2019vdj},
a procedure that is consistent from the perspective of
an expansion in the inverse number of colors \cite{Richardson:2021xiu}.
With such an estimate,
the short-range contribution 
in light nuclei is indeed comparable 
to that of the long-range mechanism 
\cite{Cirigliano:2018hja,Cirigliano:2019vdj}.
Its first inclusion in an 
{\it ab initio} calculation for an experimentally
relevant nucleus has confirmed its potential importance \cite{Wirth:2021pij}
and a strategy exists to determine its LEC directly from 
lattice QCD \cite{Davoudi:2020gxs}.

Unfortunately, the broader consequences of the Pav\'on-Phillips work 
\cite{Valderrama:2014vra}
do not appear to be fully appreciated in the community.
Moreover, the implications of an 
improved counting of $4\pi$s for the main observation in Steve's third paper 
\cite{Weinberg:1992yk}
--- the suppression of 
few-body reaction kernels,
which are the EFT version of ``meson-exchange currents'' ---
have not been explored.

\section{Delta Isobar}
\label{Delta}

I learned about the importance of the Delta isobar
while preparing for my qualifying exam
and, later, reading Gerry \cite{Brown:1972ga}
and Karl, Rupert, and Charlotte Elster \cite{Machleidt:1987hj}.
The Delta's prominence at relatively small energies comes from the 
Delta-nucleon mass difference
$\Delta\equiv m_\Delta-m_N\ll M_{\rm QCD}$.
If we do not want to limit the convergence of the EFT too severely,
$\Delta$ should be treated as a low-energy scale
and the Delta kept as an explicit degree of freedom.

Steve was pleased to hear that 
Carlos and I had already started 
including the Delta isobar when
he reported that the local group made this suggestion
during his visit to the University of Maryland in the early '90s.
There existed an immense literature on the difficulties and ambiguities
of a spin-3/2 field, but none are relevant in the nonrelativistic
expansion.
In the original submission of our first paper \cite{Ordonez:1992xp} 
we presented the two-nucleon potential up to N$^{3}$LO with the Delta.
It was an expansion in $Q/M_{\rm QCD}$ with $Q\sim m_\pi \sim \Delta$.
Our referee was supportive except for the explicit Delta.
One heard such an objection frequently in the early days,
as many chiral perturbation theorists 
seemed to think of ChPT less as a specific EFT
than as a special representation of QCD at vanishing momentum
and pion mass --- unable
to handle an expansion parameter like $\Delta/M_{\rm QCD}$
which does not vanish in these limits, 
even if it is small in the real world. 
I was in a hurry to have the paper published before going into the postdoc 
job market, so
rather than arguing we replaced the Delta components 
of the two-nucleon force by
the leading Deltaless three-nucleon force.
We did consider the Delta explicitly in the potential we fitted to data 
later \cite{Ordonez:1993tn,Ordonez:1995rz},
where it entered starting with 
the box, crossed box, and triangle TPE diagrams of rough size
\begin{equation}
\text{Deltaful two-pion exchange} \sim \frac{4\pi}{m_N M_{NN}}
\frac{Q^3}{\Delta M_{\rm QCD}^2}
\,.
\end{equation}
This is N$^{2}$LO just as
\begin{equation}
\text{Deltaless two-pion exchange} \sim \frac{4\pi}{m_N M_{NN}}
\frac{Q^2}{M_{\rm QCD}^2}
\end{equation}
if $\Delta$ is considered a low-energy scale,
but it enters only at N$^{3}$LO for $\Delta$ taken as a high-energy scale.
Integrating out the Delta, as done by
subsequent generations of chiral-potential builders,
jeopardizes the convergence pattern of Chiral EFT.

At about the same time,
Elizabeth Jenkins and Aneesh Manohar demonstrated the importance of the Delta in
ChPT \cite{Jenkins:1991es}, with the
systematic expansion in $\Delta/M_{\rm QCD}$
exploited in subsequent papers such as Ref. \cite{Butler:1992ci}.
Later, when
studied in further detail \cite{Hemmert:1997ye}, 
this expansion was rebranded the ``small-scale expansion'' 
as a nod to chiral perturbation theorists.
Perhaps thanks to this move, the Delta was eventually accepted in ChPT,
a milestone being the work by
Ulf and Nadia Fettes 
on pion-nucleon scattering 
below the Delta peak \cite{Fettes:2000bb}.
Around the Delta peak, power counting must be
modified to account for the cancellation
between energy and $\Delta$ in $s$-channel Delta propagation. 
This was done first for Compton scattering 
\cite{Pascalutsa:2002pi} taking $m_\pi \sim \Delta^2/M_{\rm QCD}$ 
and later for pion-nucleon scattering \cite{Long:2009wq}
with the more standard $m_\pi \sim \Delta$.
Beyond the Delta peak,
the Roper resonance becomes important \cite{Beane:2002ud,Long:2011rt}.

These developments of course only reinforced the motivation
to include the Delta explicitly in the nuclear potential and reaction kernels.
Over the years I had many screaming matches with Vijay Pandharipande,
particularly when he visited Caltech 
in the late '90s and insisted on the importance of the
small hole in a deuteron wavefunction ``donut''.
But one thing we could agree on was the Delta.
Vijay was convinced the Delta was crucial, particularly 
through the 
contribution (NLO in Friar's counting) of the Fujita-Miyazawa three-nucleon 
force
\cite{Fujita:1957zz}.
One cannot but admire Vijay's legendary drive when he transmitted notes 
--- a chapter from a book he was writing --- to Daniel
while he was away from home pursuing treatment at a hospital.
Daniel and I translated Vijay's point into EFT language while we were at an
INT workshop in October 2004. 
In what I have been told was Vijay's last paper \cite{Pandharipande:2005sx},
we showed in pretty concrete terms,
I thought, how omitting the low-lying Delta leads to relatively
large errors in Chiral EFT: to relate pion-nucleon scattering to 
the TPE three-nucleon force, one extrapolates in energy by 
at least $m_\pi$, thus generating errors 
${\cal O}(m_\pi^2/\Delta^2)$ in the three-nucleon force 
when the Delta is not included explicitly.
Of course these errors can be partly corrected two orders down the expansion
but including all contributions at such a high order is,
well, a tall order.

The successes of Pionless EFT \cite{Hammer:2019poc} 
question whether pion propagation is as
important for nuclear physics as previously believed.
But of course including the pion explicitly should increase the EFT range,
and in Chiral EFT there is no fundamental difficulty in accounting
for the Delta as well. 
Still, despite the evidence for the Delta's importance
from pre-EFT nuclear physics,
from the first chiral potentials, and from ChPT,
only recently has the Delta been taken seriously again in chiral potentials. 
As expected, it improves the description from
light nuclei \cite{Piarulli:2016vel,Piarulli:2017dwd} to nuclear matter 
\cite{Ekstrom:2017koy,Jiang:2020the}. 
Unfortunately,
Delta effects are not well studied in renormalizable Chiral EFT.

\section{Conclusion}
\label{conc}

Hopefully these reminiscences give a glimpse of the many contributions
to the development of news ideas which are often not apparent in
the paper trail. 
The conceptual evolution of nuclear EFTs is particularly hard to follow 
since only a very selected type of paper has percolated
through the nuclear community --- one where the notions
of renormalization, naturalness, and power counting for observables
play second fiddle to potentials and kernels derived from an unquestioned 
recipe and then fitted to data at high orders.
Only now are $\chi^2$ paranoia, Cohen syndrome, and Delta denialism
losing their grip,
see {\it e.g.} Ref. \cite{Tews:2020hgp}.

In retrospect, perhaps it was this that 
Gerry was warning us about back in the early '90s.
The root of the failure of Steve's
original recipe
is the nonperturbative renormalization of
singular interactions \cite{Beane:2000wh,PavonValderrama:2007nu},
in particular the tensor force \cite{Nogga:2005hy}
that comes from the pions' derivative couplings and, thus, 
from spontaneous chiral-symmetry breaking.
In phenomenological models (for example, Ref. \cite{Machleidt:1987hj})
the rho meson is assumed
to have a tensor coupling which is adjusted to largely cancel 
OPE's tensor force at short distances.
The need for such a cancellation had 
in fact been considered indirect evidence for the existence 
of the rho meson 
\cite{Schwinger:1942}.
In Chiral EFT the cancellation is accomplished by LECs, not
all of which obey 
NDA --- probably what Mannque means by 
``not {\it naturally} suppressed'' in his contribution
to this volume \cite{Rho:2021alh}.
It is not surprising in hindsight that the leading-order
potential without the appropriate
interactions is considered so bad by practitioners that most 
papers only show results from higher orders.
In contrast, the ordering of the short-range interactions that control the 
tensor force 
does follow from naturalness in the nonperturbative context --- just not 
from a perturbative expansion that does not 
in any case apply to the momenta of interest.

The research set in motion by the original papers 
\cite{Weinberg:1990rz,Rho:1990cf,Weinberg:1991um,Ordonez:1992xp,Weinberg:1992yk}
has achieved the longstanding goal of renormalization in nuclear 
structure and reactions, highlighting
along the way the roles played by few-body forces and the Delta isobar
--- not to mention sparking the creation of Pionless and Halo/Cluster EFTs.
This is nothing to sneeze at: it heals the scission between nuclear and particle
physics inflicted by the failure of renormalization
in the pion theories of the '50s.
But the task of constructing a consistent Chiral EFT is far from complete,
as renormalization is a necessary but insufficient constraint
on power counting.
For more on where we are in this process --- but without the anecdotes ---
see Refs. \cite{vanKolck:2020llt,vanKolck:2020plz}.

\section*{Acknowledgments}
I thank Alejandro Kievsky for prodding me to
collect my reminiscences here,
and Silas Beane and Daniel Phillips for helping me recollect them.
I am also thankful to Alejandro and Daniel for comments on the manuscript.
This material is based upon work supported in part 
by the U.S. Department of Energy, Office of Science, Office of Nuclear Physics, 
under award DE-FG02-04ER41338.

\end{document}